\def\binom#1#2{\left(\begin{array}{c}#1\\#2\end{array}\right)}
\def\h{{1\over 2}}
\def\m{{\bf m}}

\documentstyle[12pt]{article}
\begin{document}
\baselineskip 22pt

{\ }\qquad\qquad \hskip 4.3in DAMTP/96-43

{\ }\qquad\qquad \hskip 4.3in hep-th/9604101

\vspace{.2in}

\begin{center} {\LARGE SEPARATION OF VARIABLES
AND VACUUM STRUCTURE OF  ${\cal N} = 2$ SUSY QCD}
\\ \baselineskip 13pt{\ }
{\ }\\
Tomasz Brzezi\'nski\footnote{Research
supported by the EPSRC grant GR/K02244} \\
 Department of Applied Mathematics \& Theoretical Physics\\
University of Cambridge, Cambridge CB3 9EW\\
\end{center}
\begin{center}
April 1996
\end{center}

\vspace{10pt}
\begin{quote}\baselineskip 13pt
\noindent{\bf Abstract} We show how the method of separation of
variables can be used to construct integrable models corresponding to
curves describing vacuum structure of four-dimensional ${\cal N} = 2$
SUSY Yang-Mills 
theories. We use this technique to construct  models corresponding
to $SU(N)$ Yang-Mills theory 
with $N_f<2N$   matter hypermultiplets
by generalising the periodic Toda lattice. We also show that some special
cases of massive $SU(3)$ gauge theory can be equivalently described by
the
generalisations of the Goryachev-Chaplygin top obtained via separation
of variables. 

\bigskip

\end{quote}
\baselineskip 19pt

1. In  \cite{SeiWit:mon}, Seiberg
and Witten described an intimate relationship between hyperelliptic curves
and exact effective actions for ${\cal N} = 2$ $SU(2)$ gauge theories
in four dimensions. Their
results were soon generalised to 
$SU(N)$ gauge theories \cite{KleLer:...} and the
corresponding curves were found \cite{HanOz:mod}. For the $SU(N)$
gauge theory coupled to $N_f<2N$ matter hypermultiplets the curves are
given by\footnote{We use the conventions in which
for a sequence $\alpha_k,\alpha_{k+1},\ldots$, $\prod_{i=k}^m \alpha_i
= 1$ and $\sum_{i=k}^m\alpha_i =0$ if $m<k$.} 
\begin{equation}
y^2 = \left(x^N - \sum_{i=0}^{N-2} u_ix^i +{\Lambda_{N_f}^{2N-N_f}\over
4}\sum_{i=0}^{N_f-N} 
x^{N_f-N-i}t^{(N_f)}_i(\m)\right)^2 -
\Lambda_{N_f}^{2N-N_f}\prod_{i=1}^{N_f}(x+m_i),  
\label{nf>nc}
\end{equation}
Here $u_i$, $i=0,\ldots , N-2$  are
gauge-invariant order parameters, 
$m_i$, $i=1, \ldots , N_f$  are masses of the particles and $\Lambda_{N_f}$
 are dynamically generated scales ($\Lambda_{N_f} = \Lambda$ for any
$N_f\geq N$). Furthermore for any
integer $l$, $0\leq l\leq N_f$, $t^{(l)}_k(\m)$ are
symmetric polynomials in $m_i$ defined by
\begin{equation}
t^{(l)}_k(\m) = \sum_{i_1<i_2<\ldots <i_k} m_{i_1}\cdots m_{i_k}, \qquad
1\leq i_r\leq l.
\label{tk.eq}
\end{equation}

Later it
was  realised that the Seiberg-Witten theory can be reformulated
in terms of classical integrable Hamiltonian systems corresponding to
the elliptic Whitham 
hierarchy \cite{GorKri:int}.  The hyperelliptic curves of ${\cal N} = 2$
supersymmetric gauge theories are then interpreted as the spectral
curves of the Lax matrix $L(u)$ of the corresponding integrable
models. For example in 
\cite{Mar:exa}  the curve describing $SU(3)$ theory
with two massless 
particles  was found to correspond to the Goryachev-Chaplygin top. In
general the relevant 
systems turn out to be the specific limits of the Hitchin
models \cite{DonWit:sup} such as 
elliptic Calogero models \cite{Mar:int} or $SL(2)$-spin chains
\cite{GorMar:not}.  In the present paper we
suggest the use of separation of variables to explicitly construct the
required 
integrable systems. In particular we identify the curves describing
${\cal N} = 2$ SUSY QCD with three colours and even 
number of flavours $N_f$ and pairwise coinciding masses, with
separated equations of motion of 
generalisations of the Goryachev-Chaplygin top 
 constructed
recently in \cite{Sco:sep} via separation of variables. We also 
propose a generalisation of the periodic Toda lattice whose separated
equations of motion describe the vacuum structure of the
four-dimensional ${\cal N} =2$ SUSY $SU(N)$ Yang-Mills theory coupled
to $N_f<2N$ massive matter hypermultiplets.\medskip

2. We begin by showing how separation of variables can provide a
constructive procedure of building 
integrable models corresponding to  complex curves
that describe the vacuum structure of ${\cal N} = 2$ supersymmetric
gauge theories. 
Assume that we have a classical Hamiltonian system
integrable in the Liouville-Arnold sense with $n$ degrees of
freedom. Let $H_i$, $i=1,...,n$, be its first integrals in
involution. Then, at least locally,  there exist canonical variables 
$(p_i, q_i)$ which allow for separation of the Hamilton-Jacobi
equation thus leading to a system of $n$-equations
$\Phi_i(p_i,q_i;H_1,\ldots, H_n) = 0$. Each such equation describes a
curve in variables $(p_i,q_i)$. In most cases the
functions $\Phi_i$ coincide, i.e. $\Phi_i = \Phi$, $i=1,\ldots, n$,  and thus
the separated equations describe 
identical curves. The problem of solving the system reduces to
solving the equation $\Phi(p_i,q_i; H_1, \ldots, H_n)=0$. Having
solved the integrable system by separation of variables one can
construct new integrable systems by adding new terms to $\Phi$. One
starts with the separation coordinates $(p_i, q_i)$, and considers
a system described by separated equations $\Phi(p_i,q_i; H_1,\ldots, H_n) +
\Psi(p_i, q_i)= 0$, where $\Psi$ is an arbitrary function which does
not depend on $H_1,\ldots ,H_n$. One then writes $H_1,\ldots ,H_n$
in terms of original dynamical variables in which the
system was defined in the first place and thus
obtains a new integrable system which generalises the one one has
started with. In this way one can obtain a hierarchy or a family of
integrable models which separate in the same coordinates. Although the
function $\Psi$ may be arbitrary it is not always easy to find functions
that lead to physically interesting models. This method of
constructing new integrable models via separation of variables was
considered in \cite{Mac:int} in the case of Neumann model and is
fully developed in \cite{Sco:sep}.

From the point of view of
${\cal N} = 2$ supersymmetric gauge theories, one views $p_i, q_i$ (or some functions
of $p_i$ and $q_i$)  as complex
variables, and the equation $\Phi(p_i,q_i; H_1,\ldots, H_n) +
\Psi(p_i, q_i)= 0$ as a definition of a complex curve. One can then adjust
function $\Psi$ in 
such a way that the resulting curve corresponds to a given gauge
field theory and then, by converting to original variables, one can
find corresponding integrable dynamical system. For example, as will
be shown in next two sections, knowing
the curve and the model corresponding to the pure gauge theory, one
can construct a model corresponding to the gauge theory coupled to
matter. 

Let us note that the
above interpretation of hyperelliptic curves of ${\cal N} = 2$ SUSY gauge
theories in terms of separated Hamilton-Jacobi equations agrees with
that of \cite{GorKri:int} in which the curves are identified with
spectral curves of the Lax operator $L(u)$. In the modern approach to
separation of variables \cite{Skl:sep} via functional Bethe Ansatz one
starts with the eigenvalue problem of the Lax operator, $L(u)\Omega(u) =
p(u) \Omega(u)$. For many models, the poles $q_i$ of the
Baker-Akhiezer function $\Omega(u)$ Poisson commute with each other
and together with the corresponding eigenvalues $p_i = p(q_i)$ (or
some functions of $p_i$)
provide the set of separation variables.  It is clear that since for each
$i$, $p_i$ is an eigenvalue of $L(q_i)$, $(p_i,q_i)$ lie on the
spectral curve of the Lax operator, i.e. 
$\Phi(p_i,q_i)\equiv \det(p_i - L(q_i)) = 0$. This last equation gives
the separated equations of motion since the coefficients of the
characteristic polynomial of 
$L(q_i)$ depend only on the Hamiltonians $H_1,\ldots, H_n$. Thus the
spectral curve  
of $L(u)$ plays the role of a generating function of separated
Hamilton-Jacobi equations.\medskip

3. As a first  illustration of the
general procedure of interpreting hyperelliptic curves in terms of
separated Hamilton-Jacobi equations of integrable models we consider 
the $SU(3)$ curves (\ref{nf>nc}) with $N_f =2n$, $n=0,1,2$, and $m_{i+n} = 
m_i$, $i=1,\ldots, n$, and we show that  the corresponding
integrable 
system is a generalisation of the Goryachev-Chaplygin top.

The Goryachev-Chaplygin top is constructed from the variables $x_i$,
$J_i$, $i=1,2,3$ whose Poisson brackets obey the following relations
$$
\{J_i,J_j\} = \epsilon_{ijk} J_k, \quad \{J_i,x_j\}=
\epsilon_{ijk}x_k, \quad \{x_i,x_j\} = 0.
$$
The Hamiltonian of the system is
$$
H = {1\over 2}(J_1^2 +J_2^2 +4J_3^2) - bx_1,
$$
where $b$ is a free parameter. If the motion of the model is  subjected to the
constraints 
$$
x_1^2 +x_2^2+x_3^2 = 1, \qquad x_1J_1 +x_2J_2 +x_3J_3 = 0,
$$
then the induced Poison bracket is non-degenerate  and there is an
additional integral of motion
$$
G = 2J_3(J_1^2+J_2^2) + 2bx_3J_1,
$$ thus making the system completely integrable. The
Goryachev-Chaplygin top admits the separation coordinates 
$$
q_1 = J_3 +\sqrt{J^2}, \qquad q_2 = J_3 -\sqrt{J^2},
$$ 
with the conjugate momenta $p_1,p_2$ given by $\cos(2p_i) = -x_1
+x_3J_1/q_i$.  The
separated equations are
\begin{equation}
2b\cos (2p_i)+q^2_i = 2H +{G\over q_i},
\label{separated.chap}
\end{equation} 
where $H$ and $G$ are now interpreted as separation constants. The
left hand side of (\ref{separated.chap}) may  be interpreted as new
first integrals $H_i$. Thus equivalently we can define the
Goryachev-Chaplygin top by giving the $H_1$, $H_2$ as stated.

To recover curves (\ref{nf>nc}) we follow
the procedure described above and consider the generalisation of
the Goryachev-Chaplygin top  given by  the
separated equations \cite{Sco:sep}
\begin{equation}
q_i^2 +{1\over 4} A^2_n \sum_{k=0}^{2n-3} q_i^{2n-4-k}
t^{(2n)}_k(\m) + A_{n} q_i^{-1} \prod_{k=1}^n (q_i+m_k)\cos (2p_i) = 2H
+{G\over q_i} , 
\label{separated.gen}
\end{equation}
where $i=1,2$. For given $n$, the
generalisation of the 
Goryachev-Chaplygin top (\ref{separated.gen}) depends on $n$
mass parameters $m_1,\ldots ,m_n$ and the scale $A_n$. The polynomials
$t^{(2n)}_k$ are given by (\ref{tk.eq}) with $m_{n+i} = m_i$, $i=1,\ldots
,n$. We recover the
original Goryachev-Chaplygin top when $n=1$, $m_1=0$ and $A_1 =
2b$. By setting $x =q_i$ and 
$$
y= \pm iA_n\prod_{k=1}^n (q_i+m_k)\sin (2p_i),
$$
we obtain that the equations (\ref{separated.gen}) describe the curves
$$
y^2 = \left(x^3 -2Hx -G +{A^2_n\over
4}\sum_{i=0}^{2n-3} x^{2n-3-i}t^{(2n)}_i(\m)\right)^2 -
A_n^{2}\prod_{i=1}^{2n}(x+m_i), 
$$
which for $n=0,1,2$ are
equivalent to (\ref{nf>nc})  provided we make the 
identifications:
$$
H =\h u_1, \quad G = u_0, \quad A_0 =\Lambda_0^3, \quad A_1 = \Lambda_2^2,
\quad A_2 = \Lambda.
$$
We can express $H$ and $G$ in (\ref{separated.gen}) in terms
of the
natural coordinates $J_i, x_i$ and 
consider $H$ as a physical Hamiltonian of this new integrable
system. Following \cite{Sco:sep} we define
functions
$$
{\cal J}^{(r)} = {q_1^{r+1}-q_2^{r+1}\over q_1-q_2}
$$
for any integer $r$. In terms of the original variables $J_i$ the
functions ${\cal J}^{(r)}$ read
$$
{\cal J}^{(m)} =\sum_{0\leq k\leq
m/2}\binom{m-k}{k}(2J_3)^{m-2k}(J_1^2+J_2^2)^{k}, 
\qquad 
{\cal J}^{(-m)} = (-1)^m(J_1^2+J_2^2)^{-m+1}{\cal J}^{(m-2)},
$$
for any non-negative integer $m$. Using these functions we can write
Hamiltonian $H$ of the system described by separated equations
(\ref{separated.gen}) as
\begin{eqnarray*}
2H & = & J_1^2+J_2^2+4J_3^2 +{A^2_n\over
4}\sum_{k=0}^{2n-3}{\cal J}^{(2n-k-4)}t^{(2n)}_k(\m)\\
&&  + A_n\sum_{k=0}^n(-{\cal
J}^{(n-k-1)}{x_1} +{\cal J}^{(n-k-2)}J_1x_3)t^{(n)}_k(\m).
\end{eqnarray*}
The other constant of motion is
\begin{eqnarray*}
G & = & (2J_3 +{A^2_n\over
4}\sum_{k=0}^{2n-3}{\cal J}^{(2n-k-5)}t^{(2n)}_k(\m)\\
&& + A_n\sum_{k=0}^n(-{\cal
J}^{(n-k-2)}{x_1} +{\cal
J}^{(n-k-3)}J_1x_3)t^{(n)}_k(m))(J_1^2+J_2^2) .
\end{eqnarray*}
For given $n$,  the Hamiltonian $H$ and the other integral $G$ have
the following structure 
$$
H = {1\over 2} (J_1^2+J_2^2+4J_3^2) + P_n(x_i,J_i)
+{A_n\over 2}{J_1x_3\over J_1^2+J_2^2}\prod_{i=1}^nm_i,
$$
$$
G =
2J_3(J_1^2+J_2^2) +Q_{n+1}(x_i,J_i) -A_n({x_1}  +{J_1x_3J_3\over
J_1^2+J_2^2})\prod_{i=1}^nm_i 
$$
where $P_n$, $Q_n$ are polynomials of degree at most $n$. Explicitly for all
the cases relevant to the ${\cal N} = 2$ supersymmetric $SU(3)$ gauge theory 
 the polynomials $P_n$, $Q_n$ read
$$
P_0 =0, \qquad P_1 = -A_1{x_1\over 2}, \qquad P_2 = {A_2\over
2}(J_1x_3 -2J_3x_1) -{A_2(m_1+m_2)\over
2}x_1 + {A^2_2\over 8},
$$
$$
Q_1 = 0, \qquad Q_2 = A_1J_1x_3, \qquad Q_3 =-2A_2x_1(J_1^2+J_2^2) +
A_2(m_1+m_2)J_1x_3 + (m_1+m_2){A^2_2\over 2}.
$$
\medskip

4. Now we proceed to construct models
corresponding to curves that describe the vacuum structure of massive
$SU(N)$ theories with arbitrary  $N_f<2N$, as given by
(\ref{nf>nc}). It is known that for $N_f =0$ 
the corresponding 
model is the periodic Toda lattice. Therefore we seek suitable
generalisations of the Toda lattice whose separated equations of
motion could lead to all the curves given by (\ref{nf>nc}).

Recall that the periodic Toda lattice is given by the Hamiltonian
$$
H = {1\over 2}\sum_{i=1}^N\pi_i^2 + \sum_{i=1}^N e^{x_i-x_{i+1}},
$$
where $(\pi_i,x_i)$, $i=1,\ldots, N$ are canonical variables and
$x_{N+1} = x_1$. Without the loss of generality we can assume that the
total momentum $\sum_i\pi_i$, which is a constant of motion,  vanishes. 
In \cite{KacMoe:som} the separation coordinates $q_1,\ldots ,q_{N-1}$
for the Toda lattice 
were constructed as eigenvalues of the matrix $L_1$ obtained from the
$N\times N$ Lax-matrix 
$$
L = \pmatrix{\pi_1& e^{\h(x_1-x_2)}&0&\ldots &0&0&e^{\h(x_N-x_1)} \cr
 e^{\h(x_1-x_2)}&\pi_2& e^{\h(x_2-x_3)}&\ldots&0&0&0\cr
\ldots&\ldots&\ldots&\ldots&\ldots&\ldots&\ldots\cr
0&0&0&\ldots& e^{\h(x_{N-2}-x_{N-1})}&\pi_{N-1}& e^{\h(x_{N-1}-x_N)}\cr
e^{\h(x_N-x_1)}&0&0&\ldots &0&e^{\h(x_{N-1}-x_N)}&\pi_N}
$$
by deleting first row and column. The
separated equations are
\begin{equation}
2\cosh(p_i) - q_i^N = -\sum_{k=0}^{N-2}u_kq_i^k,
\label{toda.separated}
\end{equation}
$i=1,\ldots, N-1$. Here $u_0,\ldots 
u_{N-3}, u_{N-2} = H$ are  constants of motion and $p_i$ are momenta conjugate
to $q_i$ given by 
$p_i = \log|e^{x_1-x_2}\det(q_i-L_2)|$, where $L_2$ is the matrix
obtained from $L$ by deleting first two rows and columns.
 We remark in passing that
equations (\ref{toda.separated}) are equivalent to the equation $\det
(v-L(u)) =0$ for the spectral parameter dependent $N\times N$ Lax
matrix $L(u)$ (see e.g. \cite{GorMar:not}) provided we put $v = q_i$,
$u= e^{p_i}$. 

Following the general procedure described in Section 2 we consider an
integrable system constructed from the canonical variables $p_i,q_i$,
$i=1,\ldots, N-1$ defined above
and the separated equations
\begin{equation}
\h(P_{n_1}(q_i)e^{ p_i} + Q_{n_2}(q_i) e^{- p_i}) -(q_i^N
+{\Lambda^{2N-N_f}_{N_f}\over 
4}\sum_{k=0}^{N_f-N} 
q_i^{N_f-N-k}t^{(N_f)}_k(\m)) = -\sum_{k=0}^{N-2}u_kq_i^k,
\label{gen.gen}
\end{equation}
$i=1,\ldots, N-1$. Here $P_{n_1}$, $Q_{n_2}$ are polynomials such that
$P_{n_1}(x)Q_{n_2}(x) = \Lambda^{2N-N_f}_{N_f}\prod_{k=1}^{N_f}(x+m_k)$, 
and $u_0,\ldots, u_{N-2}$ are first integrals of
motion (or, equivalently, separation constants). 

The corresponding hyperelliptic curve is obtained by setting 
\begin{equation}
y =\pm\h(P_{n_1}(q_i)e^{ p_i} - Q_{n_2}(q_i) e^{- p_i}) ,
\qquad x =q_i,
\label{gen.y}
\end{equation}
and is identical with (\ref{nf>nc}).

Therefore any  classical integrable Hamiltonian system given by
separated equations (\ref{gen.gen}) provides a generalisation of the
Toda lattice which describes a vacuum structure of general ${\cal N}=2$
SUSY QCD. Now we would like to study the structure of this model from
the Hamiltonian point of view. In particular we would like to express
the model in terms of the natural variables $\pi_i, x_i$.  Following
(\ref{toda.separated}) 
we choose $u_{N-2} = H$ to be the Hamiltonian of the
system. Using properties of the Vandermonde determinant one easily
finds the explicit form of the Hamiltonian
\begin{eqnarray}
H &=& \sum_{i=1}^{N-1}{-\h(P_{n_1}(q_i)e^{ p_i} + Q_{n_2}(q_i)
e^{- p_i}) +q_i^N 
\over{\prod_{j\neq i}(q_i-q_j)}}\nonumber \\ 
&& + {\Lambda^{2N-N_f}_{N_f}\over  
4}(\delta_{N_f,2(N-1)} +\delta_{N_f,2N-1}(\sum_{i=1}^{N-1}q_i
+\sum_{k=1}^{N_f}m_k)). 
\label{h.gen}
\end{eqnarray}
Although 
the explicit expressions for $q_i$ in terms of $\pi_i$ and $x_i$
cannot be obtained the expressions of the form 
$$
\sum_{i=1}^{N-1} {F(q_i)\prod_{j\neq i}(q_i-q_j)^{-1}},
$$
where $F$ is any rational function 
can
be found in terms of original variables since they involve only
symmetric polynomials $t_k^{(N-1)}({\bf q})$ in the eigenvalues
$q_1,\ldots q_{N-1}$
of matrix $L_1$. The polynomials $t_k^{(N-1)}({\bf q})$  can be easily
read off from the characteristic polynomial of $L_1$,
e.g. $t_1^{(N-1)}({\bf q}) =  
{\rm tr} L_1$ and $t_{N-1}^{(N-1)}({\bf q})=\det L_1$ etc. This suffices to
find $H$ as  
a function of $x_i, \pi_i$ since, by definition of the $p_i$,
$$
P_{n_1}(q_i)e^{ p_i} + Q_{n_2}(q_i) e^{- p_i} =
P_{n_1}(q_i)e^{x_1-x_2}|\det(q_i-L_2)| +
{Q_{n_2}(q_i)e^{x_2-x_1}\over|\det(q_i-L_2)|}
$$ 
is a rational function of $q_i$. Furthermore, with no loss of
generality we can assume that $n_1\geq n_2$ and take 
$$
P_{n_1}(q) = \Lambda^{N-n_1}_{N_f}\prod_{k=1}^{n_1}(q+m_k), \qquad Q_{n_2}(q)
= \Lambda^{N-n_2}_{N_f}\prod_{k=n_1+1}^{N_f}(q+m_k).
$$
Then, the  Hamiltonian $H$
(\ref{h.gen}) has the following form
\begin{eqnarray*}
H & = & {1\over 2}\sum_{i=1}^N\pi_i^2 + \sum_{i=2}^{N-1} e^{x_i-x_{i+1}}
+{\Lambda^{2N-N_f}_{N_f}\over  
4}(\delta_{N_f,2(N-1)} +\delta_{N_f,2N-1}(
\sum_{k=1}^{N_f}m_k -\pi_1))\\
&& + {\Lambda_{N_f}^{N-n_1}\over 2}(\prod_{k=1}^{n_1}(\pi_2 +m_k)e^{x_1-x_2}
+\Lambda^{n_1-n_2}_{N_f}\prod_{k=n_1+1}^{N_f}(\pi_N +m_k)e^{x_N-x_1}
+G^{(N)}_{N_f}) ,
\end{eqnarray*}
where $G^{(N)}_{N_f}$ are  polynomials in $\pi_2, \ldots ,\pi_{N}$ of degree
$n_1-2$. For a given $N$, $G^{(N)}_{N_f}$ are defined for any $N_f
=0,\ldots, 2N-1$ and are obtained as contributions to $H$ coming from
the terms of the form
$$
\sum_{i=1}^{N-1}{q_i^ke^{ p_i}
\over{\prod_{j\neq i}(q_i-q_j)}}+\pi_2^ke^{x_1-x_2}, \qquad  
\sum_{i=1}^{N-1}{q_i^ke^{- p_i}
\over{\prod_{j\neq i}(q_i-q_j)}}+\pi_N^ke^{x_N-x_1}.
$$
As an explicit  example we  take
$n_1, n_2$  such that $n_1-n_2$ is either 0 or 1. 
The first seven  $G^{(N)}_{N_f}$ come out as:
$$
G^{(N)}_0 = G^{(N)}_1 = G^{(N)}_2 = 0, \qquad 
G^{(N)}_3 =  e^{x_1-x_3}, \qquad G^{(N)}_4 = e^{x_{N-1}-x_1} + e^{x_1-x_3},
$$
$$
G^{(N)}_5 = \Lambda_5 e^{x_{N-1}-x_1} + (2\pi_2 +\pi_3 +\sum_{k=1}^3
m_k)e^{x_1-x_3}, 
$$
$$
G^{(N)}_6 = (2\pi_N +\pi_{N-1} +\sum_{k=4}^6 m_k) e^{x_{N-1}-x_1} +
(2\pi_2 +\pi_3 +\sum_{k=1}^3 m_k)e^{x_1-x_3}, 
$$
and
\begin{eqnarray*}
G^{(N)}_7 & = &  \Lambda_7(2\pi_N +\pi_{N-1} +\sum_{k=5}^7 m_k)
e^{x_{N-1}-x_1} + 
(\sum_{k<l}^4m_km_l + (2\pi_2 +\pi_3)\sum_{k=1}^4 m_k\\
&& +3\pi_2^2
+2\pi_2 \pi_3 +\pi_3^2 )e^{x_1-x_3} +e^{x_1+x_2-2x_3} +e^{x_1-x_4}
\end{eqnarray*}
The $G^{(N)}_{N_f}$ functions listed above  suffice to describe
all the Hamiltonians of integrable models corresponding to $SU(3)$ and
$SU(4)$ gauge theories.

Notice that the generalised Goryachev-Chaplygin Hamiltonian derived in
Section 3 can be viewed as a special case of the $N=3$ Hamiltonian
(\ref{h.gen}) provided we make suitable identifications of $q_i, p_i$
and $P_{n_1}$, $Q_{n_2}$.

We would like to conclude the paper by indicating the possibility of
yet another 
description of the curve (\ref{nf>nc}) which makes explicit use of
interpretation of $y$ and $x$ as functions on the phase space of the
integrable Hamiltonian system  given by (\ref{gen.gen}). The
Hamiltonians $u_k$ give rise to Hamiltonian vector fields parametrised
by the `times' $t_0,\ldots, t_{N-2}$ and, for any function $f$ on the
phase space,  given by
$
{\partial\over{\partial t_k}} f = \{ u_k,f\}.
$
Using definition (\ref{gen.y}) of $y$ and $x$ and separated equations
(\ref{gen.gen}) we can  express the curve
(\ref{nf>nc}) as
$$
y = \pm \sum_{k=0}^{N-2}x^k{\partial x\over \partial t_k}.
$$
We think that this simple equation can prove useful in analysis and
interpretation of the canonical form on
the hyperelliptic curve, which constitutes the
second half of the Seiberg-Witten theory. The analysis of the
canonical one-form from this point of view
is currently being
carried out and we hope to present the results of this investigation
soon.

\end{document}